# Feature Selection Strategies for Classifying High Dimensional Astronomical Data Sets

Ciro Donalek, S. G. Djorgovski, Ashish A. Mahabal,
Matthew J. Graham, Andrew J. Drake
California Institute of Technology
Pasadena (CA), USA
donalek@astro.caltech.edu

Thomas J. Fuchs, Michael J. Turmon
Jet Propulsion Laboratory
California Institute of Technology
Pasadena (CA), USA

Arun Kumar A., N. Sajeeth Philip
St. Thomas College
Kerala, India

Michael Ting-Chang Yang
Graduate Institute of Astronomy, NCU
Taiwan, Taiwan

Giuseppe Longo
Università degli Studi Federico II
Napoli, Italy

*Abstract* — The amount of collected data in many scientific fields is increasing, all of them requiring a common task: extract knowledge from massive, multi parametric data sets, as rapidly and efficiently possible. This is especially true in astronomy where synoptic sky surveys are enabling new research frontiers in the time domain astronomy and posing several new object classification challenges in multi dimensional spaces; given the high number of parameters available for each object, feature selection is quickly becoming a crucial task in analyzing astronomical data sets. Using data sets extracted from the ongoing Catalina Real-Time Transient Surveys (CRTS) and the Kepler Mission we illustrate a variety of feature selection strategies used to identify the subsets that give the most information and the results achieved applying these techniques to three major astronomical problems.

*Keywords* — astroinformatics; machine learning; feature selection; CRTS

## I. Introduction

In the last decade astronomy, as well as many other scientific fields, has experienced a huge growth in volume, complexity and even quality of data, both from actual measurements and from numerical simulations. In astronomy, such growth has been caused mainly by the availability of new, panoramic digital detectors which have opened the way to the new generation of synoptic sky surveys [1], producing an amount of data to be ingested and analysed daily which doubles every 12-18 months. A second factor which is contributing to this data avalanche is the availability of heterogeneous data sets collected over the years by ground based and space borne instruments, now available through web-services [2]. In spite of the differences present in each specific area, most of the basic tasks to be performed are in common, such as, for instance, statistical analysis, clustering, high dimension visualization; in many cases this multi parametric data need to be processed in quasi real-time fashion in order to extract knowledge as rapidly and efficiently possible.

An automated reliable and robust classification of transient and variable sources is of a critical importance for the effective follow-up of the present and future synoptic sky surveys. When discovered, all transients look the same (see Fig. 1) and the main question that arises is "how do we decide which are the most interesting objects worthy of follow-ups with expensive facilities"? Most systems today rely on a delayed human judgment in decision making and follow-up of events and this "manual" approach will simply not scale to the next generation of surveys.

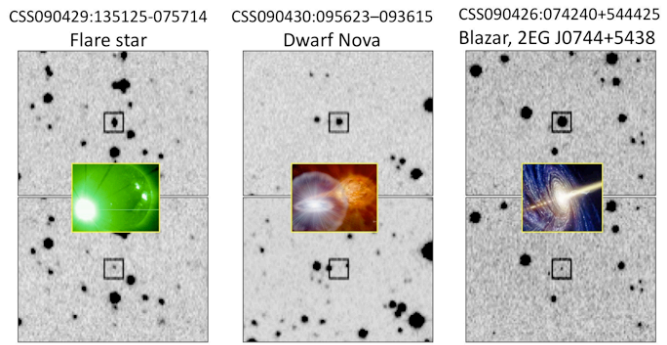

Fig. 1. Top row shows objects which appear much brighter that night, relative to the baseline images obtained earlier (bottom row). On this basis alone, the three transients are physically indistinguishable. Subsequent follow-ups show them to be three vastly different types of phenomena.

A major problem associated with pattern recognition in high-dimensional data sets is the curse of dimensionality; this can be addressed by selecting only a subset of features that are rich in discriminatory power with respect to the classification problem at hand. Feature selection is preferable to feature transformation (e.g., Principal Component Analysis) when the meaning of the features is important and one of the modeling goal is to find meaningful relationship between the parameters in order to understand the physical nature of the problem. Roughly speaking, in the assumption that any observable quantity or observing parameters can be expressed by a numerical measure, the astronomical Parameter Space is an $N$-dimensional numerical space and every observation becomes a $N$-dimensional feature vector and we are interested in finding the $m$ dimensions, with $m \ll N$, that best describe our data set. Reducing the number of features is also in line with the goal of avoiding overfitting to the specific training data set and of designing classifiers that result in good generalization performance [3].

In this paper we show the implementation of a machine learning approach to three major astrophysical problems where the reduction of the dimensionality of the input parameter space enables better results.

## II. DATASETS

The catalogs used in the experiments contain objects, from Catalina Real-Time Transient Survey (CRTS) [4] and the Kepler Mission [5]. Each astronomical object in represented through its light curve that can be sparse and uneven sampled, with tremendous variations also in errors, number of points, missing values, etc., making comparison between them as well as training the classifiers difficult. To overcome this problem we extracted a set of ~60 statistical and morphological descriptors (see, e.g., [6][7][8]) using the Caltech Time Series Characterization Service (CTSCS, see Fig. 2) [9]; vectors of such features derived from the light curve of known classes of objects were used as training and test sets in our classification system.

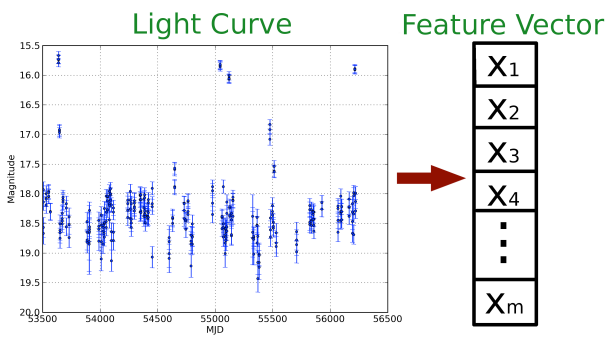

Fig. 2. Through the CTSCS a set of statistical descriptors are extracted from each light curve forming a feature vector. The plot shows a light curve of a Cataclysmic Variable from CRTS.

### A. Catalina Real-Time Transient Survey

Catalina Real-Time Transient Survey (CRTS) [4] is systematically exploring and characterizing the faint, variable sky. It covers the total area of ~ 33,000 deg$^2$, down to ~ 19 - 21 mag per exposure, with time baselines from 10 min to 8 years, and growing; there are now typically ~ 300 - 400 exposures per pointing, and coadded images reach deeper than ~ 23 mag.

The survey has detected ~ 7,500 unique, high-amplitude transients, including at least 1,800 supernovae, at least a 1,000 CVs (the majority of them previously uncatalogued), over 2,500 of blazars / OVV AGN, hundreds of flare stars, etc. The survey has a complete open data policy: all transients are published immediately electronically, with no proprietary period at all. Furthermore over 500 millions light curves have been released.

### B. Kepler Data

The Kepler Mission [5] was launched to survey a large number of stars for planets, and then determine their properties like sizes, masses, densities, reflectivity, etc. Kepler completed its main mission in November 2012 when it entered an extended phase and currently it is in a safe mode. The data it collected continues to be analyzed and new discoveries continue to be made. The speciality of the Kepler observations are the dense lightcurves required for various planet detection methods and because of the relatively small area it covers (115 square degrees). These are in complete contrast with the rather sparse lightcurves that individual objects from sky surveys like CRTS possess (CRTS coveres nearly 3000 times Kepler's area).

## III. FEATURE SELECTION STRATEGIES

Feature selection algorithms can be roughly grouped in to two categories: filter methods and wrapper methods. In the filter methods the feature selection is independent of the classifier used to evaluate the results; they rely on general characteristics of the data to evaluate and to select the feature subsets without involving a specific learning algorithm. Wrapper methods use the performance of the selected algorithm to evaluate each candidate feature subset, searching for features better fit for the chosen learning algorithm, but they can be significantly slower than filter methods if the learning algorithm takes a long time to run.

In this work we studied five different feature selection strategies and evaluated the results through classifiers trained using only the subsets found by these algorithm versus the results obtained using all the features available for each problem.

### A. Fast Relief Algorithm

Fast Relief Algorithm (aka ReliefF) [10][11] is a simple yet efficient procedure to estimate the quality of attributes. It is usually applied in data pre-processing as a feature subset selection method. The key idea of the ReliefF Algorithm is to estimate the quality of attributes according to how well their values distinguish between instances.

Relief algorithms in general compute two quantities: the average weight vector $W$ and the threshold $\tau$. $W$ is calculated for each feature by finding two nearest neighbors: one from the same class (nearest hit) and the other from different class (nearest miss); it shows how relevant a feature is in representing a class while the actual selection of the feature is based on the value of $\tau$. This class of algorithms are found to be robust also in presence of noisy data.

## B. Fisher Discriminant Ratio (FDR)

The Fisher Discriminant Ratio (FDR) [12] can be used to rank a number of features with respect to their class-discriminatory power and can be independent of the type of the underlying class distribution.

Let $\mu_1$ and $\mu_2$ be the means of class 1 and class 2 respectively and $\sigma_1$ and $\sigma_2$ the corresponding variances, the FDR is defined as:

$$FDR = (\mu_1-\mu_2)^2 / (\sigma_1^2 / \sigma_2^2)^2$$

Features having large differences between the means of the classes and small variances in each class will have a greater FDR value and will be ranked higher than the others. This method can be applied only to binary classification problems.

## C. Correlation-based Feature Selection (CFS)

Correlation-based Feature Selection (CFS) [13] is a wrapper method which selects features that have low redundancy and is strongly predictive of a class.

It is based on the hypothesis that features that are strongly predictive of a class are highly correlated with the class, yet uncorrelated with each other. The CFS algorithm basically uses symmetrical uncertainty which is a measure of feature – class correlation. The symmetrical uncertainty also gives a measure of goodness of the selected feature subset.

A forward best first mechanism is used to find the subset of features with high class correlation, with a stopping criterion such that five consecutive fully expanded subsets show no improvement over the current best subset.

## D. Fast Correlation Based Filter (FCBF)

Fast Correlation Based Filter (FCBF) [13] is a supervised filter based feature selection algorithm. This method is similar to the CFS algorithm in the sense that it also uses feature-class correlations and symmetrical uncertainty as measures of goodness.

The FCBF algorithm probes for features that have predominant correlation with the class or become predominant after removing its redundant peers. The algorithm works in two stages. In the first stage it ranks features based on the value of symmetrical uncertainty and in the second stage removes features that are not predominant. This method is designed to be computationally efficient with very high dimensional data.

## E. Multi Class Feature Selection (MCFS)

Multi Class Feature Selection (MCFS) [13] is basically an unsupervised feature selection method based on the spectral analysis of the data. The specialty of this algorithm is that, even though inherently unsupervised, it can be used in supervised as well as semi-supervised modes. The algorithm uses spectral analysis of data and $L_1$ regularized models for subset selection to select feature subsets which preserves the multi cluster/class structure of the data.

The method first constructs a graph to identify the $p$ nearest neighbors for each data point and obtains a weight matrix $W$. A Laplacian graph is obtained such that $L = D - W$ where $D$ is a diagonal matrix whose entries are column sums of $W$. This is then solved as a generalized eigenvalue problem to find the feature subset by minimizing a fit error using least angle regression. Another peculiarity of the algorithm is that the user can specify the number of best features to be returned.

## IV. CLASSIFIER DESIGN

We employed different classifiers in the selected feature space to assess the performances of the feature selection routines. In this paper we demonstrate that feature selection strategies actually help in reduce the dimensionality of the problem without a loss in performance.

The performance of the classifiers were rated based on the following three criteria: *completeness*, the percentage of objects of a given class correctly classified as such; *contamination*, the percentage of objects of a given class, incorrectly classifed as belonging to another class; *loss*, fraction of misclassified data.

The performance of a given classifier in terms of its error rate was measured against the data set using a 10-fold Cross-Validation approach in which the original samples are randomly partitioned into 10 subsamples; each time a single subsample is retained as test, and the remaining are used as training data. This process is then repeated 10 times with each of the 10 subsamples used exactly only once as test set.

## A. Ensembles of K-nearest-neighbor (KNN)

The *K*-nearest-neighbor (KNN) [14] is a very simple method for classification. Despite the simplicity of the algorithm, it can performs very well and, more important is often used as a benchmark method.

Suppose that there are $c$ classes, for a new pattern $x$ we compute the distances to all the other training examples and select a subset $S_k$ consisting of the $K$ closest values. Let $k_i$ be the frequency of the $i$-th class in $S_k$ Then we assign $x$ to the class $C_m$ with the greatest frequency: $k_m >= k_i$ for $i=1,,..,c$. If there is a tie, the winning class is chosen randomly [15]. When using KNN, the curse of dimensionality basically means that the distance become meaningless in the high dimensional space with all vectors basically equidistant from the centroids. Feature extraction help to avoid this effect.

## B. Ensembles of Decision Trees

Decision Trees (DT, aka Classification Trees) [16] are one of the most used data mining, non-linear classifier. In a decision tree the feature space is split into unique regions, corresponding to the classes. Each internal node denotes a test on an attribute, each branch represents the outcome of the test and each leaf holds a class label.

In our tests, the DTs were trained first using all the features available, and then on the substes generated by the feature selectio algorithm described above. Each tree was built using the Gini Diversity Index (gdi) as criterion for choosing the split; the splitting stops when there is no further gain that can be made. All the results are shown below.

## V. EXPERIMENTS

The problems covered in this paper are inherently multi-class problems, but given the high number of classes and the heterogeneity of the data, we found that an hierarchical approach led to better results [17]. We used some astrophysically motivated major features to separate different groups of classes (see Fig. 3), and then proceeding down the classification hierarchy each node uses those classifiers that are demonstrated to work best for that particular task.

Binary classification is an increasingly common task in Astronomy [18][19]. In this paper we have considered two binary classification problems using CRTS data and a three-classes classification problem using Kepler data.

TABLE I.

| SN vs "ALL THE OTHERS" (see Table IV for the complete parameters description) | | |
|---|---|---|
| *Feature Selection Strategy* | *KNN Loss* | *DT Loss* |
| None (all parameters selected) | 30% | 18% |
| ReliefF (6 parameters selected: $x1, x2, x19, x17, x15, x7$) | 22% | 15% |
| CFS (3 parameters selected: $x2, x8, x13$) | 24% | 17% |
| FCBF (3 parameters selected: $x2, x8, x13$) | 24% | 17% |
| MCFS (4 parameters selected: $x9, x13, x14, x16$) | 32% | 19% |
| FDR (6 parameters selected: $x15, x5, x8, x14, x16, x17$) | 22% | 16% |

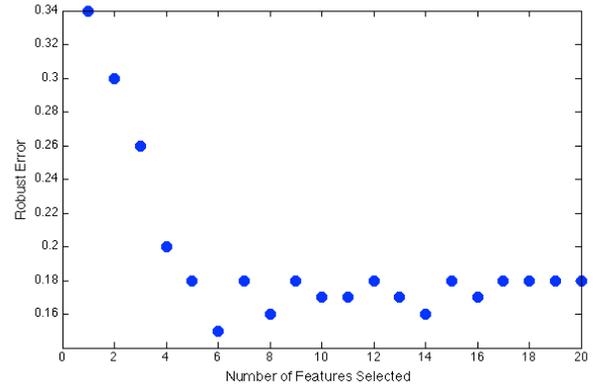

Fig. 4. The plot shows how the misclassification error change based on the number of features used to train the DT. The features were ranked using the ReliefF algorithm. Best result was achieved using the six most important features found by the algorithm.

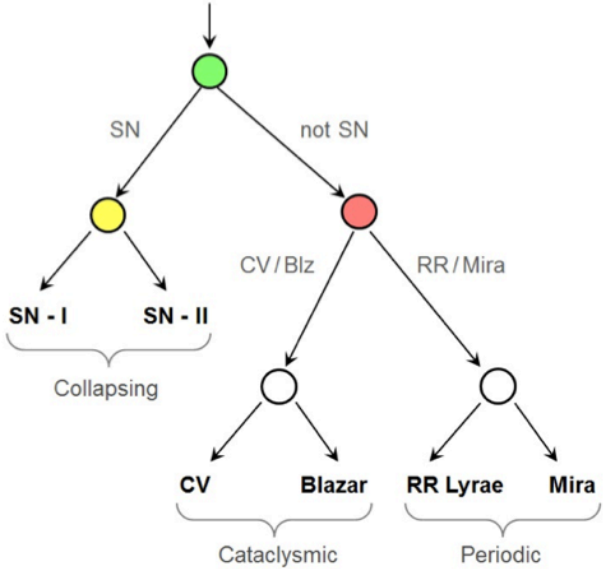

Fig. 3. Hierarchical approach: some astrophysically motivated major features are used to separate different groups of classes. Proceeding down the classification hierarchy each node uses those classifiers that are demonstrated to work best for that particular task.

### A. Supernova Classification

The goal of this binary classification problem is to reliably assign each object in one of two mutually exclusive classes: Supernovae or not-Supernovae. To perform this task, we extracted from CRTS light curves belonging to six different classes of objects: Supernovae, Cataclysmic Variables, Blazars, other AGNs, RR Lyrae and Flare Stars. Then, using the CTSCS, twenty features were extracted from each light curve and used as training set.

Results obtained used different combinations of feature selection strategies and classifiers are shown in Table I. Fig. 4 shows the misclassification error versus the number of features used to train the DT; best result was achieved training the DT with the first six parameters ranked by the ReliefF algorithm, with a significant gain in accuracy for both KNN and DT classifiers.

### B. WuMa vs RR Lyrae in CRTS

Eclipsing binaries (W UMa) are the main contaminant in studies using RR Lyrae as tracers of Galactic structures [20] and therefore being able to distinguish between them would be crucial for that study. To study this problem we extracted CRTS light curves for 482 RR Lyrae and 463 W UMa and used the CTSCS to extract 60 periodic and non-periodic features for each object.

Best results in terms of classification were achieved using the five higher ranked parameters according the ReliefF algorithm. Table II show completeness and contamination for both classes, achieved using all the parameters and the selected subset. Not only using the best five parameters according to the ReliefF algorithm decreased dramatically the computational time but also led to a higher completeness and lower contamination. It is interesting to note that the parameters automatically selected by this procedure, essentially represent the period-amplitude relationship illustrated in Fig. 5 which is used to differentiate between subclasses of RR Lyrae. Other approaches using information-theory methods and symbolic regression led to comparable results [7].

TABLE II.

| W UMa vs RR Lyrae using DT (see Table IV for the complete parameters description) |||
|---|---|---|
| **Feature Selection Strategy: None** | **Completeness (DT)** | **Contamination (DT)** |
| W UMa | 93% | 7% |
| RR Lyrae | 94% | 6% |
| | | |
| **Feature Selection Strategy: ReliefF (5 parameters selected: x22, x10, x14, x19, x15)** | **Completeness** | **Contamination** |
| W UMa | 97% | 3% |
| RR Lyrae | 96% | 4% |

TABLE III.

| Kepler Data (see Table IV for the complete parameters description) || |
|---|---|---|
| **Feature Selection Strategy** | **KNN Loss** | **DT Loss** |
| None (all parameters selected,) | 16% | 15% |
| ReliefF (6 parameters selected: $x12, x2, x21, x14, x10, x7$) | 15% | 13% |
| CFS (7 parameters selected: x1, $x7, x9, x15, x16, x21$) | 17% | 16% |
| FCBF (4 parameters selected: $x7, x9, x16, x17$) | 19% | 18% |
| MCFS (5 parameters selected: $x7, x9, x15, x16, x21$) | 17% | 18% |

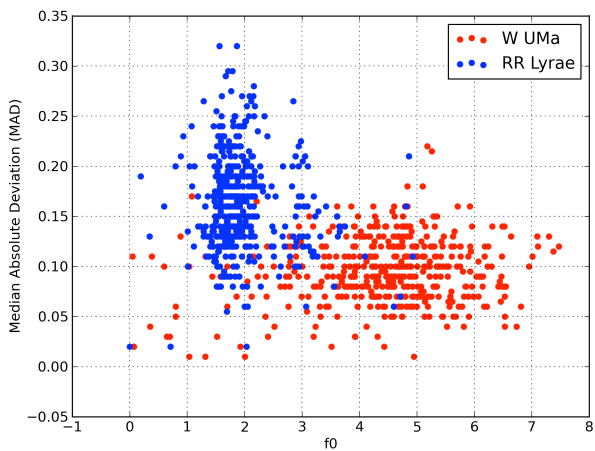

Fig. 5. It is interesting to note that the best two parameters automatically selected by the ReliefF algorithm (f0, mad), essentially represent the period-amplitude relationship [7] and give a good separation between the two classes.

## C. Classifying objects in Kepler Data

As of July 2013 Kepler has found well over 100 confirmed exoplanets, and there are over 3000 additional candidates. The subsets we have considered here include ~20 thousands objects that Kepler has observed: Red Giants, Eclipsing Binaries, and other Kepler Objects of Interest (KOIs). It is this latter group that consists of stars with planets around them, but also contains many false positives which are also of interest to us. Our investigations include: comparing light curve parameters for dense versus sparse lightcurves, understanding which subset of light curve parameters are better suited to separate the three classes, and to try and understand the nature of the false positives if possible. That's why a reliable classification of these objects is needed.

For this experiment we ran four feature selection strategies to a 3-class problem, and then test the subsets generated by them using both KNN and DT. Table III show the results achieved; the best overall classification was reached using a DT and the six best parameters found by the ReliefF algorithm.

## VI. CONCLUSIONS

In this paper we addressed the need to extract relevant subsets of features from multiparametric data sets in order to reduce the dimensionality of the input space and achieve better results decreasing the complexity of the problem. The three experiments conducted all show the advantage of employing feature extraction routines rather than using all the features available both in terms of computational time and overall classification rate.

In the Supernova classification problem, we found that all the feature selection strategies led to a better classification rate rather than using all the parameters; in particular, the best result was found using the six higher-ranked parameters according to the ReliefF algorithm with a significant gain in accuracy for both KNN and DT classifiers.

Also in the second binary classification problem addressed we found that using only five parameters out of the ~60 available led to a significantly higher completeness and lower contamination for both classes; moreover, we were also able to retrieve some meaningful relationship in a totally unsupervised way (eg, the period-amplitude relationship in RR Lyrae).

The third experiment was based on Kepler data and the goal was discriminating between three different classes of objects: eclipsing binaries, red giants and KOI. In this case we trained our classifiers to recognize all three classes and we found that best results were achieved training a DT with the most six relevant discriminant features according to the ReliefF algorithm. Also in this case, the loss is lower than the one achieved using all the 21 parameters available.

Feature selection algorithms are expected to become increasingly common and useful in astronomy where the large number of features available for each object make the traditional analysis difficult to perform. We expect that the methods described above will be deployed in a near future for the analysis of data from CRTS and other sky surveys.

ACKNOWLEDGMENTS

S.G.D., C.D., A.A.M., and M.J.G acknowledge a partial support from the NSF grants AST-0834235 and IIS-1118041, and the NASA grant 08-AISR08-0085. Some of the work reported here benefited from the discussions during a study and the workshops organized by the Keck Institute for Space Studies at Caltech.

TABLE IV.

**List of Features Extracted Using the CTSCS**
**Parameters available for the SN classification: x1 to x20; Kepler object classification: x1 to x21; binary classification WUMa vs RR Lyrae: x1 to x59.**

| Name | Variable | Description |
|---|---|---|
| id | x0 | Object ID |
| amplitude | x1 | Half the difference between the minimum and maximum magnitudes |
| beyond1std | x2 | Percentage of points beyond one standard deviation from the weighted mean |
| fprm20 | x3 | Ratio of flux percentiles: (60th-40th) over (95th-5th) |
| fprm35 | x4 | Ratio of flux percentiles: (67.5th-32.5th) over (95th-5th) |
| fprm50 | x5 | Ratio of flux percentiles: (75th-25th) over (95th-5th) |
| fprm65 | x6 | Ratio of flux percentiles: (82.5th-17.5th) over (95th-5th) |
| fprm80 | x7 | Ratio of flux percentiles: (90th-10th) over (95th-5th) |
| linear_trend | x8 | Slope of a linear fit to the light curve |
| max_slope | x9 | Maximum absolute flux slope between two consecutive observations |
| mad | x10 | Median discrepancy of the fluxes from the median flux |
| mbrp | x11 | Percentage of fluxes within 10% of the amplitude from the median |
| pair_slope_trend | x12 | Percentage of the last 30 pairs of consecutive flux measurements that have a positive slope |
| percent_amplitude | x13 | Largest percentage difference between either the maximum or minimum flux and the median |
| pdfp | x14 | Percent Different Flux Percentile: ratio of (95th-5th) flux percentile over median flux |
| skew | x15 | Skew of the magnitudes |
| s_kurt | x16 | Kurtosis of the magnitudes |
| std | x17 | Standard deviation of the light curve |
| rcorbor | x18 | Fraction of magnitudes 1.5 magnitudes below the median |
| mag_r | x19 | Indicates whether the object spends most of its time above or below the median [21] |
| stetson_j | x20 | Welch-Stetson J variability index with an exponential weighting scheme |
| stetson_k | x21 | Welch-Stetson K variability index |
| f0, f1, f2 | x22-x24 | The three prime frequencies from the frequency analysis statistics in [8] |
| Frequency parameters | x25-x49 | The first four harmonics (amplitude and phase) for f0, f1, f2 and other frequency statistics as described in [8] |
| Lomb-Scargle peaks | x50-x59 | The periods and false-peak detection probabilities of the top 5 peaks in the Lomb-Scargle periodogram of the light curve |
| class | x60 | Object class |